# Characterization and on-sky demonstration of an integrated photonic spectrograph for astronomy

N. Cvetojevic,[1] J.S. Lawrence,[1,2,*] S.C. Ellis,[3] J. Bland-Hawthorn,[3] R. Haynes,[1] and A. Horton[1]

[1]*Anglo-Australian Observatory, PO Box 296, Epping, NSW 2121, Australia*
[2] *MQ Photonics Research Centre, Department of Physics and Engineering, Macquarie University, NSW 2109, Australia*
[3]*School of Physics, University of Sydney, NSW 2006, Australia*
*[*]jsl@science.mq.edu.au*

**Abstract:** We present results from the first on-sky demonstration of a prototype astronomical integrated photonic spectrograph (IPS) using the Anglo-Australian Telescope near-infrared imaging spectrometer (IRIS2) at Siding Spring Observatory to observe atmospheric molecular OH emission lines. We have succeeded in detecting upwards of 27 lines, and demonstrated the practicality of the IPS device for astronomy. Furthermore, we present a laboratory characterization of the device, which is a modified version of a commercial arrayed-waveguide grating multiplexer. We measure the spectral resolution full-width-half-maximum to be 0.75±0.05nm (giving $R = \lambda/\delta\lambda = 2100\pm150$ at 1500nm). We find the free spectral range to be 57.4±0.6nm and the peak total efficiency to be ~65%. Finally, we briefly discuss the future steps required to realize an astronomical instrument based on this technology concept.

## 1. Introduction

The next generation of major ground-based optical and near-infrared astronomical telescopes, Extremely Large Telescopes, will have aperture sizes from 20–42 meters in diameter, far greater than existing telescopes. This has a major impact on required imaging and/or spectroscopic instrumentation, as for traditional designs the size of the instrument grows in proportion to the telescope aperture and the cost of the instrument increases with the telescope aperture squared or faster [1]. Further demands on instrument size and cost arise from the desire to obtain spectra from thousands of locations in the telescope focal plane simultaneously.

The emerging field of astrophotonics offers solutions to such problems [2]. Astrophotonic devices demonstrated to date include fiber Bragg gratings that suppress atmospheric infrared emission [3], and planar waveguides for beam combination in stellar interferometry [4].

Particularly relevant to massive-multiplexed imaging spectroscopy is the proposed integrated photonic spectrograph (IPS) [1]. One of the suggested IPS formats employs an arrayed-waveguide grating (AWG) structure that inputs light from a standard optical fiber and outputs a focal plane spectrum that is dispersed over a region of several millimeters, rather than the hundreds of centimeters associated with conventional spectrographs [5]. Such a device requires no moving components or reimaging optics.

AWGs have been extensively used in the telecommunications industry as optical multiplexers and de-multiplexers in dense wavelength division multiplexed systems [6,7]. As such, they are optimized for the needs of the telecommunications industry. For instance, the minimization of crosstalk by having more array channels than output channels, and by having multiple fiber-optic outputs. Many of these requirements are less stringent for an IPS used as an astronomical instrument.

The IPS prototype used in this work consists of a commercial planar-fabricated AWG that has been modified for astronomical application. In this paper we present an overview of the operation of AWGs, a description of our prototype IPS device, and the results of our laboratory characterization. Lastly, we show the first astronomical on-sky demonstration of an IPS, conducted at the Anglo-Australian Telescope site at Sidings Springs, Australia.

## 2. Arrayed-waveguide grating modifications

The standard AWG comprises an input fiber port, which in the case of our prototype is an SMF-28 single-mode fiber, which feeds into a multiplexor. The multiplexor, or free-propagation zone, disperses the light into a parallel array of closely spaced single-mode waveguides, which in turn feed a demultiplexor (another free-propagation zone). The array waveguides have incremental lengths, and hence the phased array behaves like a grating. At the central wavelength, a constant phase profile is exhibited at the end of the arrayed waveguides, an integer number of cycles out of phase along the plane perpendicular to the direction of propagation. Different wavelengths will have a different phase change in this plane. The demultiplexor allows the light to interfere, and focuses it onto the output ports that lie along its concave output facet [6].

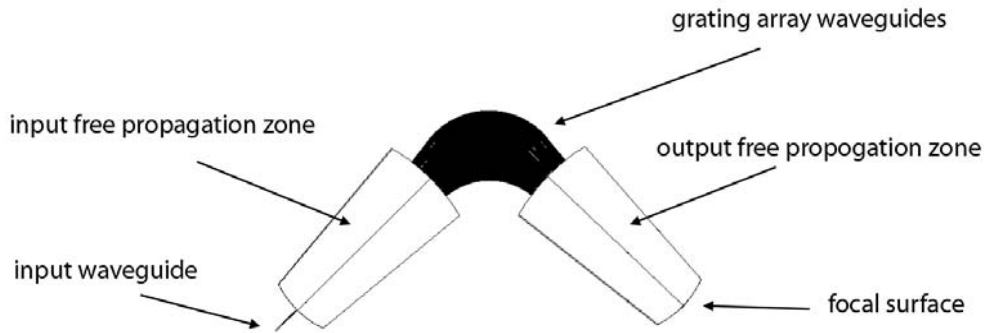

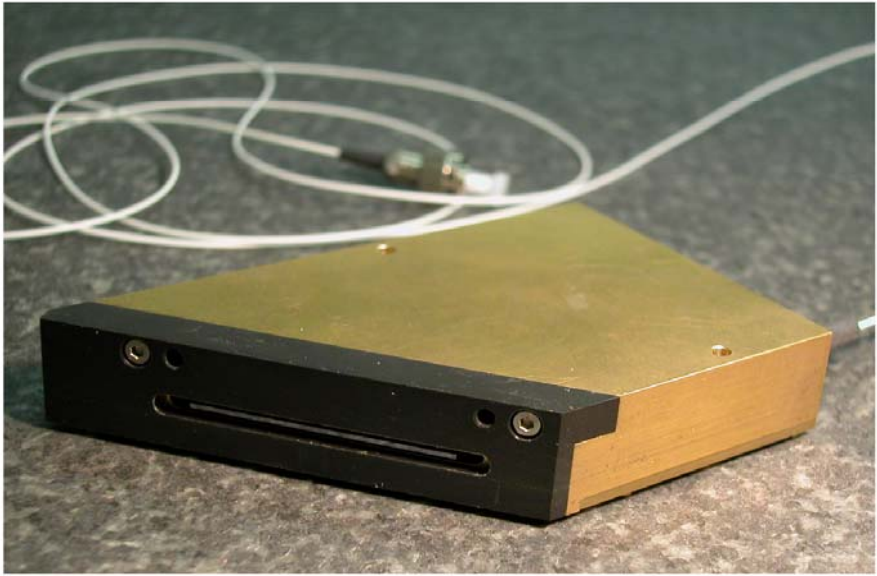

Fig. 1. Top: schematic of the AWG chip. Bottom: photograph of the IPS prototype. In this prototype device, the output array waveguides (standard in commercial devices) have been removed, the output facet has been polished back to the focal surface, and the chip packaged. The fiber input is at the rear of the image, and the polished flat output facet at the front. The open face of the chip is ~38mm in length. The active area is ~3mm across, and is situated at the centre of the polished face.

Commercially available AWGs consist of between 8 and 40 output channels, with the channel spacing typically 100GHz (~1nm) or 50GHz (~0.5nm). Recently, low-crosstalk AWGs with a channel spacing of 10GHz (0.08nm) have been demonstrated under laboratory conditions [8,9]. Most commercial AWG devices operate with a central wavelength of 1550nm, where the attenuation in telecommunication fiber is the lowest, and at high diffraction order (>20). All waveguides in an AWG are typically single-moded at the operational wavelength, as multimode propagation is less predictable and may cause more cross-talk between channels. Few-mode devices, however, have been demonstrated [10,11].

Fig. 1 shows a schematic of the AWG we have used for this work. The grating has a single input waveguide (of width 8μm) with ~22mm focal length free-propagation zones and 428 array waveguides (with a length increment of ~28μm). The device is designed to operate at a centre wavelength of 1550nm with a 100GHz output channel separation and 40 output waveguide channels.

Several modifications were made to this commercial AWG to create an IPS that can be used for astronomy. The primary hurdle was that the output consisted of a series of single-mode fibers, each carrying a wavelength channel. Since we wish to obtain a continuous spectrum at the device output these individual channels are not required. The output waveguide array was removed by slicing the chip near the output of the demux, and polishing down towards the output focusing surface, so as to remove any waveguide 'stumps' left by the cleaving process. Finally, a custom package was constructed so the IPS has one single-mode input fiber and an exposed output facet to allow coupling to an external lens or imaging array (Fig. 1).

Wang et al. (2003) [12] previously reported a similarly modified AWG device for use in optical monitoring, with a flat, non fiber-coupled, output. That device was not designed as an imaging system, but utilized a series of photodiodes to derive the channel power information at high diffraction orders.

Flat-field AWGs have been proposed using aberration theory to remove the Rowland curvature, which is inherent to this technology [13]. Our prototype does not have any such correction and is simply polished flat. Thus, we expect the IPS's focal plane to have a radius of curvature of ~22mm, defined by the focal length of the output free-propagation zone. In both the laboratory and on-sky imaging experiments, however, the curvature had little measurable affect on the resulting image, as the depth of the focal plane curvature is only ~90μm at the edges of the active region (the region over which the single on-axis input waveguide is dispersed at the focal plane of the output free-propagation zone). If the magnification of the imaging optics is greatly increased, a detector array placed directly at the focal surface of the IPS, or if the active length is increased, then the aberrations caused by the curvature may become significant.

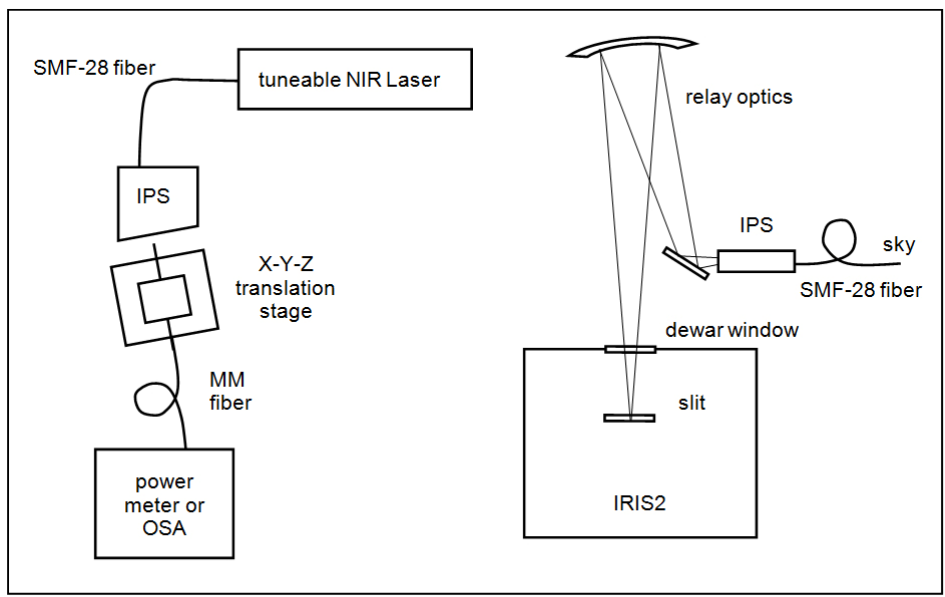

Fig. 2. Left: schematic showing the setup for the IPS laboratory characterization experiments.
Right: schematic of the on-sky experiments using the IRIS2 infrared imaging spectrometer.

## 3. Characterization of the integrated photonic spectrograph

The IPS was characterized by coupling a tunable diode laser (Santec TSL-210), which can scan between 1500nm and 1580nm at increments of 0.01nm, into its input fiber. To monitor the IPS output we used a 50μm core diameter multimode fiber on a precision x-y-z translation stage, allowing positioning within 5μm, coupled to a laser power meter or optical spectrum analyzer (OSA). A multimode fiber is used (rather than a single mode fiber) to achieve better coupling performance of the output spectrum as the core diameter is larger than the spectral point spread function (psf). The setup is shown in Fig. 2. The tunable diode laser was fixed to have a constant output power of 3mW (stable to within 1%) at the central wavelength for the duration of the experiments to acquire accurate throughput measurements. The central peak FWHM of the input laser was at or less than the OSA's maximum resolution of ~0.04nm, allowing for accurate measurements of the spatial and spectral psf.

The diffraction efficiency and throughput were measured simultaneously by translating the multimode fiber along the focal surface of the IPS at 10μm intervals with the laser operating at a fixed wavelength. Once the output spectral profile was acquired, the laser wavelength was moved by 5nm, and the process repeated. The psf measurements were conducted with the optical fiber stationary and the laser wavelength stepped through 0.05nm intervals, equating to ~1.5μm spatial resolution (for a linear dispersion of 33nm/mm).

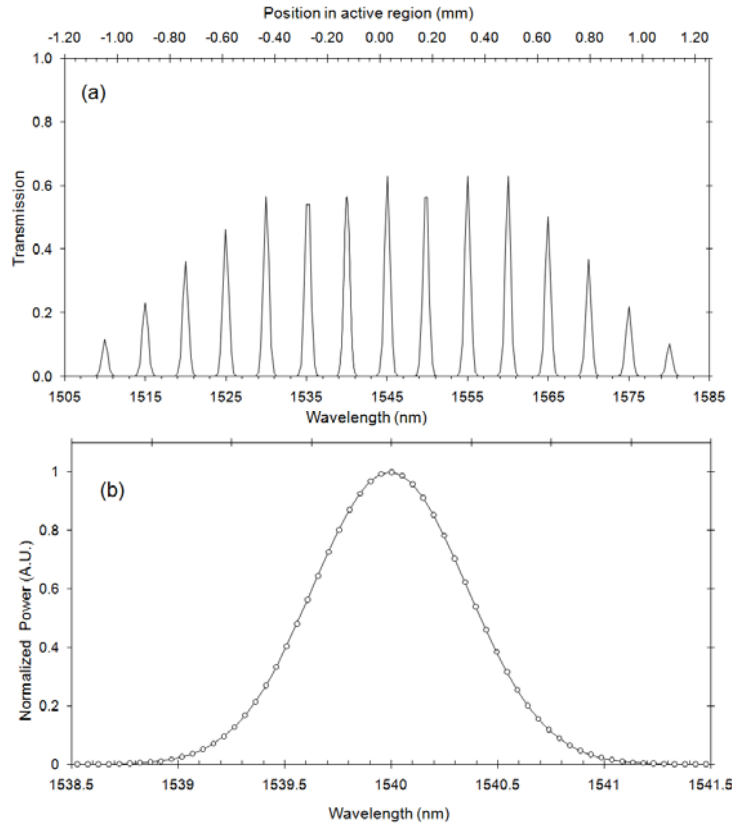

Fig. 3. (a) Total IPS efficiency across the full laser wavelength tuning range (this corresponds to ~90% of the device active area). Each peak corresponds to a series of focal plane measurements at a fixed input laser wavelength. (b) High-resolution de-convolved measurement of the output IPS spectral point spread function obtained by tuning the central laser wavelength by 0.05nm increments at a fixed fiber position in the focal plane.

Because the IPS psf was narrower than the multimode fiber's core diameter (50μm), we were effectively dithering across the function. Thus, to recover the actual output of the IPS we de-convolved our Gaussian-like measured function with the theoretical acceptance aperture of a multimode fiber. This de-convolution process is the primary determinant of uncertainties in our measurements of both the spectral psf and the spectral efficiency. By examining a range of functional forms (i.e., Butterworth, Gaussian, Moffat functions) we determine the errors to be ~7% in spectral resolution and ~8% in spectral efficiency.

## 4. Results

The total efficiency (including diffraction efficiency, throughput, and output coupling losses) of the IPS is shown in Fig. 3(a) for the full laser tuning range. The device operates principally in the 26$^{th}$ diffraction order at these wavelengths. It shows relatively high efficiency (50–60%) over a large fraction of its active area. The IPS spectral psf is shown in Fig. 3(b). We estimate the spectral resolution FWHM to be 0.75±0.05nm, corresponding to a spatial resolution of 23±1μm with a linear dispersion of 33nm/mm, and a resolving power R=$\lambda/\delta\lambda$=2100±150. Furthermore, there is little observed variation in the FWHM of the output spectral psf across the laser tuning range.

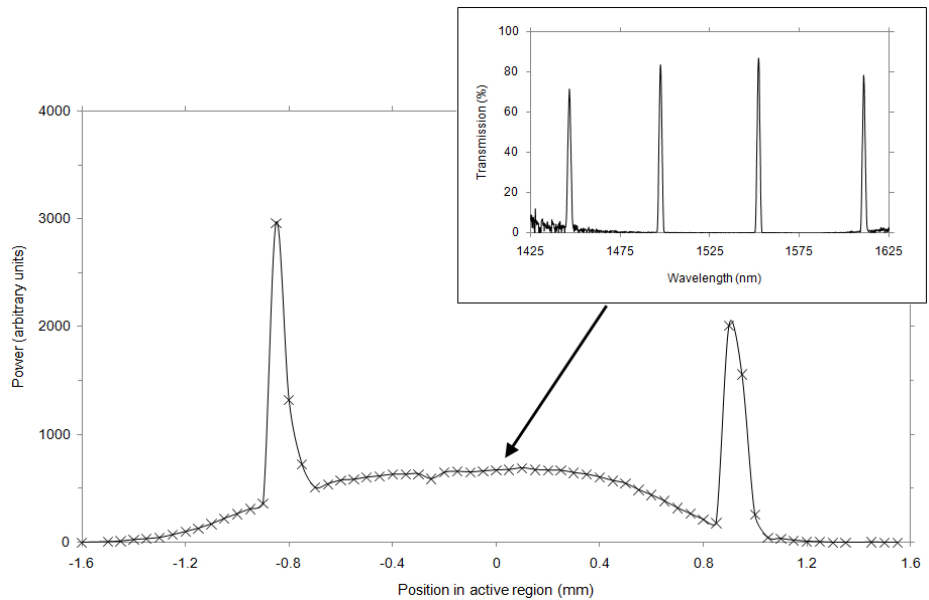

Fig. 4. Power output across the full IPS focal plane (measured by translating a fiber) with the input laser locked to 1580nm. The two peaks are adjacent orders of the same laser wavelength. Insert: Normalized spectral measurement (using the OSA) with the fiber fixed in the IPS focal plane position indicated by the arrow, demonstrating the transmission of higher diffraction orders from the wings of the laser gain profile.

The IPS free spectral range and the full spatial dimensions of the active area are illustrated in Fig. 4. Physically, the total active area is ~2.8mm. This region contains ~1.6 diffraction orders with a free spectral range of 57.4±0.6nm. The central region of Fig. 4 shows significant power is transmitted outside the laser central wavelength. By observing this background using the multimode fiber and OSA we were able to separate it into its resulting components (shown as an insert in the figure). We find that the majority of the background is due to the transmission of wavelengths at higher and lower orders through the IPS originating from the wings of the laser gain profile. Scattering of light (i.e., due to undispersed or redispersed stray light or small-scale wings of the spectral psf) around the central orders is very low (<0.1%). There is evidence of increased scattering well outside (a few times the FSR) the central operating wavelength. However, this is a high noise measurement, as the laser power is very low at these wavelengths.

The insert in Fig. 4 also shows that the transmission does not drop off dramatically with higher or lower orders, but rather has a gradual decline moving off the optimized 26$^{th}$ order.

## 5. On-sky demonstration of the IPS

On-sky data was obtained with the prototype IPS using the Anglo-Australian Telescope IRIS2 near-infrared imager and spectrograph at Siding Spring observatory during observing runs from the 17$^{th}$ to 21$^{st}$ of June, 2009. The IRIS2 instrument [14] employs a 1024×1024 pixel HAWAII-1 HgCdTe infrared detector array, which is illuminated by an f/8 to f/2.2 focal reducer. A range of standard astronomical filters for imaging are installed inside the instrument dewar, as are a slit wheel and a sapphire grism for spectroscopy. The spatial scale of the IRIS2 detector is 0.45 arcsec/pixel. The 1" slit for spectroscopy projects to 2.2 pixels on the detector, providing a resolution of R~2400.

The IPS input fiber was fixed viewing the night sky at a ~45º zenith angle. We used a relay optics assembly to re-image the IPS output focal plane onto the IRIS2 slit (see Fig. 2). This assembly gives a magnification of 3.5, resulting in a final plate scale of 1.2 pixels per IPS resolution element. Due to the lack of strong atmospheric emission lines within the narrowband (i.e., smaller than the IPS free spectral range) IRIS2 filters we used a broadband H filter (1440–1820nm) and a sapphire grism to cross-disperse the IPS output spectrum. In this mode the multiple orders of the IPS which are superimposed on the instrument entrance slit are spread out in a series of bands across the detector array.

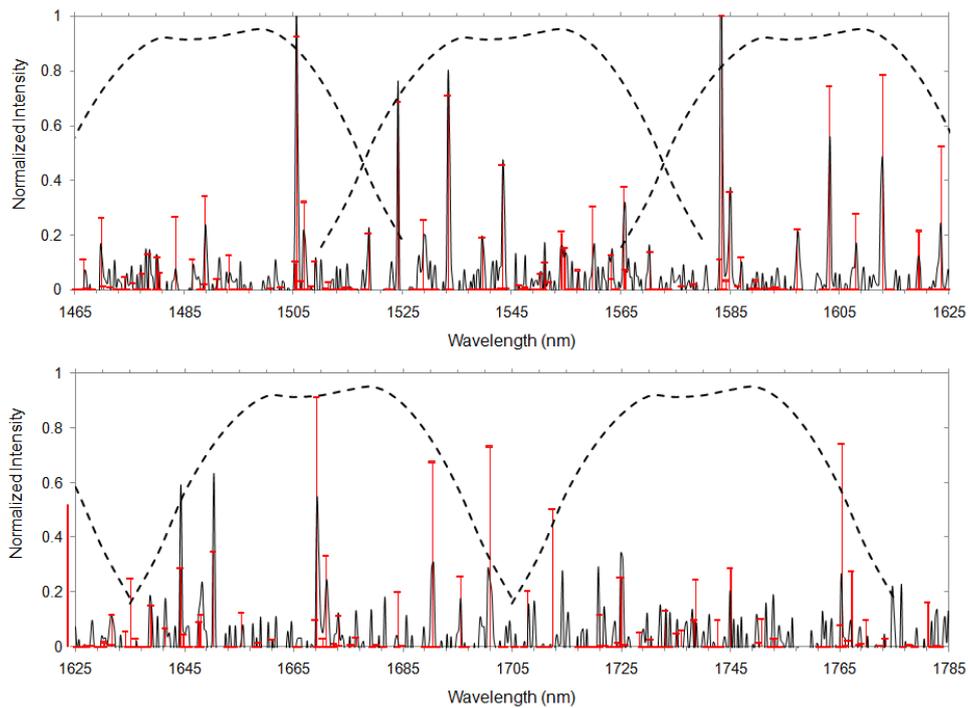

Fig. 5. The night sky OH spectrum (solid black lines) using the IRIS2 instrument as a cross-disperser for the IPS, superimposed with the theoretical [3] positions and strengths of the OH lines (red), and the diffraction efficiency envelop of the IPS for each order (dashed lines). The 23$^{rd}$ order is top left, which progresses to 27$^{th}$ order in the bottom right of the graph.

Using a series of 15min exposures over a period of 12hrs, we were able to detect the atmospheric emission spectrum as shown in Fig. 5. In this case, the spectral resolution obtained is a convolution of the IRIS2 resolution with the IPS resolution. To the authors' knowledge, this is the first continuous spectrum to be imaged using an arrayed-waveguide grating spectrometer.

The spectrum in Fig. 5 shows the detection of upwards of 27 atmospheric OH emission lines across the 380nm filter bandwidth. If we project the dispersed spectrum across the IRIS2 slit axis (i.e., giving the IPS spectral resolution) then the average FWHM across the 27 measured sky lines is consistent with the resolution (i.e., within experimental error) as measured by the laboratory tests. The intrinsic variation in the magnitude of the atmospheric OH emission lines combined with the long exposure times required (due to the small fiber etendue) precludes a determination of the IPS spectral efficiency via the on-sky test.

## 6. Conclusion

With the characterization and first on-sky demonstration of an integrated photonic spectrograph, we have shown that using an astrophotonic approach to solve the problem of the rising cost of astronomical instrumentation is not only possible, but is practical as well.

We have measured the total efficiency of our IPS prototype device to be ~60% at the central wavelength. This is significantly higher than the ~30% currently achieved with typical astronomical multi-object spectrometers. The R~2000 resolution of the device is modest by astronomical standards but is sufficient for a broad range of scientific applications, particularly if used in combination with OH suppression technology, for example, where the requirement for high spectral resolution to observe sources between atmospheric emission lines is relaxed at near-infrared wavelengths. Additionally, it is known that AWG devices can be designed to operate with much higher spectral resolution.

While we have demonstrated the feasibility of an IPS for astronomy, there is still a significant gap between this prototype device and a future astronomical instrument based on this technology. Firstly, we have obtained an output spectrum from the device using reimaging optics and cross-dispersion from a second spectrometer; a proof-of-concept using a single detector array at the output curved focal surface is still required. Secondly, the free spectral range of the prototype device is set at ~55 nm due to requirements of telecommunications applications; for astronomy this must be increased by a factor of at least six to cover a complete atmospheric waveband (ie, J, H, or K) in the near-infrared. Finally, this device employs a single-mode fiber waveguide as the input source; this leads to low coupling efficiencies for astronomical telescopes unless they are operating at close to the diffraction limit. Further work is thus required to produce devices with few-mode or multi-mode input fibers, which would allow astronomical source (rather than sky) spectra to be obtained.

There are several alternatives to the AWG technology proposed here for an astronomical IPS, including photonic echelle gratings [1] and miniature spectrographs that employ micro-optics [15]. These alternatives will be pursued in the next phase of this research. To properly exploit any of these integrated photonic devices will require massively multiplexed instruments and low read noise and low dark current infrared detectors [16].